*********************************************** 
\magnification 1200
\centerline{\bf Studies in Structure formation in theories with a
repulsive long range gravitational force}
\vskip 3cm
\centerline {Daksh Lohiya, Annu Batra, Sameer Mehra}
\centerline {Shobhit Mahajan and Amitabha Mukherjee}
\centerline {Department of Physics \& Astrophysics, University of
Delhi,}
\centerline {Delhi 110 007, India}
\centerline {email: dlohiya@ducos.ernet.in}
\vskip 3cm
\centerline {\bf Abstract}
\vskip 1cm
    This article reports on emergence of structures in a class of 
alternative theories of gravity. These theories do not have any 
horizon, flatness, initial cosmological singularity and (possibly)
quantization problems. The model is characterised by a dynamically
induced gravitational constant with a ``wrong'' sign corresponding
to
repulsive gravitation on the large scale. A non - minimal coupling
of a scalar field in the model can give rise to non - topological 
solitons in the theory. This results in domains (gravity - balls)
inside which an effective, canonical, attractive gravitational
constant 
is induced.

      We consider simulations of the formation and evolution of
such
solutions. Starting with a single gravity - ball, we consider its 
fragmentation into smaller (lower mass) balls - evolving by mutual
repulsion. After several runs, we have been able to identify two
parameters: the strength of the long range 
gravitational constant and the size
of the gravity balls, which can be used to generate appropriate
two point correlations of the distribution of these balls.

\vfil\eject

     Standard Big - Bang [SBB] theory provides a framework within
which
one could address a variety of problems in Cosmology. 
Particularly regarded as ``well accounted'' are the primordial
light element synthesis and the cosmic microwave background
radiation.
However, when applied to large scale structure formation, the
status
of SBB is not quite as respectable. More than 90\% of the
gravitating 
content of the universe is required to be non - luminous and in
an as yet unknown form. Attempts to account for large 
scale structure in the Universe by a suitable combination of hot
and cold dark matter has met with  debatable success 
[see eg. Padmanabhan 1993, for a review]. Fixing the total amount
of dark matter by reference to the closure density of the universe
and fixing the hot component by requiring potential wells to form
at large scales, one is not left with sufficient residual [cold]
dark matter to account for enough power to accommodate smaller
[galactic and cluster] scale structures. Further, the proximity
of the density of the universe to the closure density poses the 
flatness problem. Related to this problem is the difficulty in  
obtaining the homogeneous and isotropic Robertson Walker solution
dynamically. The FRW metric is a very special metric with 
measure zero in the space of anisotropic solutions.

       On the quantum theoretical side, we have further problems.
There is no viable quantum field theory of gravity. If we treat
the Einstein - Hilbert action as a fundamental quantum action, we
get a non - renormalizable theory. A promising program, originally 
proposed by Sakharov and much
exalted in the early 80's, was developed by Adler and Zee [see
eg. Adler 1982]. The idea was to generate gravitation from an 
effective action induced by quantized (renormalizable) matter 
fields on a curved spacetime background. Explicit calculations,
made for a large class of renormalizable models [Zee 1982], 
realize the prescription. While the sign of the gravitational 
constant turns out to be sensitive to the infrared details of
the theory, a large class of models give an induced 
gravitational constant with a ``wrong'' sign. In a previous
article [Lohiya 1995] we considered a non - minimally 
coupled scalar field in addition to an induced negative 
gravitational constant. The model is described by the action:
\vskip 1cm
$$ S = \int d^4x\sqrt{-g}[-\epsilon\phi^2R + 
{1\over 2}g^{\mu\nu}\phi_{,\mu}\phi_{,\nu} - V(\phi)
+ \beta_{ind}R + \Lambda_{ind} + L_m]\eqno{(1.1)}$$
\vskip 1cm
$\beta_{ind}$ and $\Lambda_{ind}$ being the induced gravitational
and cosmological constants, $L_m$ the matter action for the 
rest of the matter fields and $V(\phi)$ the effective potential
for the scalar field.
Derrick's theorem is not an impediment to the existence of 
non topological soliton solutions in such a model. For such
solutions, $\phi$ is constant inside a sphere and rapidly 
goes to zero near its surface.
Such solutions would have an effective canonical attractive
gravitation in their interior and have repulsive gravitation
outside. We have been exploring the possibility that 
such non topological domains - gravity balls (g-balls) -  
of the size of a typical halo of a galaxy [or larger], play 
an essential role in cosmology. 

       Large scale dynamics of an FRW universe with a gravitational
constant of the ``wrong sign'' is not saddled with any horizon,
flatness or initial singularity problems. We have further
shown [Batra 1995, Lohiya et al 1995] that the resulting dynamics
of a hot FRW universe can consistently account for the right 
amount of Helium synthesis. Studies on primordial metallicity
are still in progress.

      The standard theory of density perturbations gets drastically
modified in such a model. In standard cosmology one picks up a seed
perturbation and hopes to construct schemes that could follow its 
growth to the non linear regime: namely scales over which a 
typical galaxy could form. With a repulsive gravitational constant,

density perturbations do not grow at all in a standard
FRW model [Lohiya and Savita 1995]. Instead, one expects the
universe
to homogenize - density perturbations freezing to a constant
in co moving coordinates. One could follow the evolution of a large
dense cloud expanding in an FRW spacetime while fragmenting 
into smaller sections as the overall density perturbation
approaches
a constant. At some time during its evolution g - balls could
form in different pockets of such a cloud. These would then be the 
sites where canonical non - linear perturbations would grow into 
a typical galaxy. Alternatively one can entertain the 
possibility that at some epoch
in the past, a large g-ball is formed. The stability of
such a ball at any epoch would sensitively depend upon the 
effective potential of the scalar field and the bulk parameters 
of the universe, namely, the temperature and the trace of the 
stress energy tensor of matter fields. Such an unstable g - ball 
could split into smaller, stable g - balls in
a random manner. For the purpose of the present paper, we have 
considered successive fragmentations of a cloud in a random manner.
In
the alternative picture, our considerations would equally well 
describe a large g - ball, when destabilized at
any epoch, splitting into two balls in a randomly chosen direction
-
separated by a distance which would be of the order of the 
size of a typical ball. We have written a simple, user-friendly
simulation code in Pascal which we have run on PC 486 DX-2 and 
Pentium machines. The next section describes the program in its 
essential detail and in the last section we describe the result
of some 100 runs.
\vskip 1cm
{\bf Section II:}
\vskip 1cm
        A gravity ball is a non topological soliton solution 
in a theory described by eqn[1.1]. The effective gravitational
constant is negative outside the ball and has the canonical 
positive value inside. We assume that the parameters of the 
effective potential would support the existence of g - balls
as large as a typical galactic halo. The 
simulation described assumes that
at some epoch a large gravity ball forms. The centre of the ball
is chosen as the initial point of our simulation.
The energy $m_b$ of the ball bears a relation to the size $r_o$
of the ball. The relation is an option available in the code. 
The default relation used most in our simulations 
is $m_b \approx r_o^2$.
We assign an integer number m to signify 
the mass of the initial ball. Balls at any epoch are labelled
by another integer label, i, which runs from 1 to n: the number
of balls at any epoch. The program prescribes a procedure
``splitmass'' in which we randomly split the mass $m(i)$ of the
$i$ th ball, distributing it over two balls: the $i$ th and the
$n+1$ th ball. For most simulations we randomise the
mass of the final balls after a split, such that one of them
has a mass equal to 50\% to 75\% of the initial ball and 
the other ball has the balance 50\% to 25\%. 
The speed of both the final balls just 
after the split is the same as that of the parent $i$th ball before
the split. Energy and momentum conservation are thus automatically
taken care of in this procedure. The split points [ball centres]
are placed in a randomly oriented direction at a distance
proportional to $\sqrt{m[i]_{final}} + \sqrt{m[n+1]}$. The 
proportionality constant is a parameter ``splitrad'' that 
can be varied in our simulations. From the moment the balls 
split, their dynamics are governed by the following set of 
equations in co moving coordinates of the FRW metric: 
[see eg. Borner [1988] chapter 12.]:
\vskip 1cm
$${\vec v_i} = {\dot {\vec x_i}} [i = 1,......N]$$
$${\dot{\vec v_i}} + 2H(t){\vec v_i} ={\vec F_i}$$
$${\vec F_i} = \sum_{j\neq i} 
Gm_j {{\vec x_i} - {\vec x_j}\over |x_i - x_j|^3}\eqno{(2.1)}$$
\vskip 1cm

We integrate these equations numerically as a set of difference
equations. As described in the previous section, the simulation
equally well describes a highly correlated cloud which evolves
in an FRW spacetime - fragmenting into smaller sections that evolve
by mutual repulsion. 

     The instance of splitting is randomised by choosing
an arbitrary prescription that ensures that the probability of
splitting of higher mass balls at any given epoch is greater 
than that for lower mass balls. This is prescribed in a separate
procedure which again can be altered if desired. 
We ran several simulations where the instance of splitting
of a ball of given mass was not random. Instead, we 
chose a monotonic
functional dependence of the splitting time 
with mass that ensured the higher mass
balls split faster than the lower mass balls. Splitting stops
when all balls have mass $m_i = 1$. Thus the final number of balls
equals the mass assigned to the initial ball. We thus assume that 
a ball with mass $m_i = 1$ is stable. The evolution may be
continued
even after all splitting is over. 

     The simulation is displayed by standard graphics displays.
We have built in user-friendly menu 
options for pausing, changing the orientation 
of the axes, changing the scale, recording 
the configurations of all particles for thin real
time slices for replays, loading previous simulations and / or
going
back to the current simulation. The simulation continuously
displays the current status of a particular run, viz. the current
number of balls, the number of time steps, the value of the
simulation time 
and the maximum distance
over which the points in a run have spread. The
three dimensional depth is displayed by assigning colour to all
the points: red dots representing the farthest - the
colour going to blue for the nearest (to the starting plane).

    The purpose of the entire exercise is to explore the 
possibility that the g - balls would be regions where matter would
condense to give galactic structures.  The minimum distance over
which 
the balls are split is taken to be the size of a typical galactic
halo.
This fixes the distance scale for the simulation. To set the
time scale, we note that the scale factor scales as the cosmic time
$t$
for most part in the matter dominated era. The Hubble parameter
$H_o$
scales as $t^{-1}$. Taking $H_o^{-1} \approx 10^{10}$ years
we may consider the original g - ball to form well after 
matter and radiation decouple,
(say at $H^{-1} \approx 10^8$ years) and match the final
configuration 
of the simulation with current data on galaxy-galaxy correlation
functions.
The inverse of the Hubble parameter at the start 
sets the time scale for our simulation. For example, a time 
run of 100 units for the above mentioned range of $H_o$ gives a
scale
of $10^8$ years per unit. 

      We performed some preliminary runs for a chosen value of the 
gravitational
constant. It was found that, after an initial rapidly
changing profile, the distribution starts simply scaling
conformally. 
This is as expected from the linear theory of perturbations
[Lohiya, Savita 1995].
We ran several trials initially to ascertain a cutoff
distance beyond which we could ignore gravitational terms [the 
contribution to the rhs in eqn(2.1)]. This was done by running
a particular simulation and storing the system configuration
for different times. 
Intermediate configurations
were then re-loaded with different distance cutoffs.
The distance cutoff for which the final configuration does
not change much can thus be determined by trials. 
We can thus introduce a cutoff distance 
in our gravitational term - ignoring 
contributions from balls beyond that distance. This considerably
improves the speed of the program. 

      On terminating the simulation, one can opt for inbuilt
procedures that can output positions and velocities in formats 
suitable to be read off by appropriate statistical packages. We
built 
in a specific procedure that evaluates two point auto - 
correlation function. We enclose the entire simulation within 
a sphere centered around the initial starting point, the radius 
of the sphere being of the same order of magnitude as the average
distance over which our simulations spread. We ensure that any 
runaway particle gets reflected back from the surface of the 
sphere. This amounts to the assumption that any particles lost
from the sphere are compensated by incoming particles escaping from
neighbouring, similarly evolving, domains in a homogeneous, 
isotropic universe. The autocorrelation function is evaluated by 
contrasting the final distribution with a random distribution
of $m$ points within the sphere.

      The input parameters are thus the value of gravitational
constant
and the size parameter of the gravity balls. A complete
listing of the program can be made available on request. We have
also
designed some test files which test the randomisation and 
approximation procedures
for known closed orbits for canonical attractive gravity. 
For any given value of input parameters chosen
in the main program, one can run corresponding test files to
determine 
appropriate increments for time to be used in the difference
equations to
get the desired accuracy.
\vskip 1cm
{\bf III Results:}
\vskip 1cm

    The purpose of this exercise was to explore the possibility 
that a typical g - ball can be the region inside which non - linear
perturbations grow. Density perturbations $(\delta\rho/\rho)$ 
with repulsive
gravity approach a constant in time. One could fix this constant by

matching the perturbations with the 
COBE result at the recombination epoch. The large scale
distribution of galaxies could be generated by the bifurcation 
and evolution of the gravity balls. 

     Figures I describes a typical simulation snapshot.
In figures II to V we plot the two point correlation functions for
the specified ranges of distances for different values of the 
gravitational constant. In every simulation we can identify 
distance ranges over which the correlation function falls off
as $r^{-1.7}$ to $r^{-2}$. One can of course identify distance
ranges
over which the fall off is around the observed $r^{-1.8}$. 

      The distance at which the autocorrelation function is unity
is sensitive to the chosen gravitational constant. Assuming the
size of a typical galactic halo to be some 25 kpc, and recalling
this to be the distance between the g - balls at the time of 
splitting, the distance scale in our simulations is fixed. Assuming
that the simulation terminates at epoch $t_o$, 
the number of time steps for a typical 
simulation being $100$, the initial time and the time unit are 
then fixed at $10^{-2}t_o$. This would be the case if the gravity
balls
form and evolve at such an initial time. The units of time and
distance
then fix the gravitational constant. At present we do not 
have any rigorous theory of 
formation of gravity balls or their bifurcation. If these form at 
any epoch $t_i$, our simulations ends at $100t_i$. 
A careful study of parameters 
that could fix the epoch of formation of non - topological solitons
is thus necessary and is under progress.

     We find our results quite encouraging and are in the process 
of designing programs for much larger distributions that could be
used in larger and faster machines. 

\vskip 2cm

Acknowledgment: Helpful discussions with  Jim Peebles are
gratefully acknowledged.
 
\vfil\eject

\centerline{\bf References}

\item{1.} D. Lohiya [1994], ``Non - topological solitons in
non - minimally coupled scalar fields: theory and consequences''
Ast. and Sp. Sci. [to be published].

\item{2.} D. Lohiya, Savita G. [1995] ``Density perturbations 
in a theory with repulsive long range gravitational forces'',
[submitted for publication].

S. Weinberg, Gravitation and Cosmology, John Wiley \&
Sons, [1972].

\item{3.} G Borner, ``The Early Universe'', Springer Verlag,
[1988].

\item{4.} E. W. Kolb and M. S. Turner, ``The Early 
Universe'' Addison-Wesley, [1989].

\item{5.} T. Padmanabhan, in ``The Future utilization of Schmidt 
telescopes'' eds. J. Chapman et al., ASP Conf. series Vol. 84,
1993.

\item{6.} S.L. Adler,  Rev. Mod. Phys, Vol. 54, 729, [1982].

\item{7.} A. Batra, M. Sc. 
dissertation, Dept. of Physics \& Astrophysics, 
University of Delhi 1995, unpublished; Poster paper in
Int. Conf. Grav. Cosm., 1995, Pune.

\item{8.} A. Zee in ``Unity of Forces in the Universe'', Vol II,   
ed A. Zee, p 1062, World Scientific [1982]; Phys. Rev. Lett., 42,
417 
[1979], Phys. Rev. Lett., 44, 703, [1980].  
\vskip 2cm

Figure Captions:
  
     Figure I: A typical snapshot of a simulation after the
development freezes in co - moving coordinates.

     Figures II - V: Two point autocorrelation function for
different simulations.

Note: The figures will be posted on request, either  as PostScript
files or by FAX.
\bye